\documentclass[12pt,preprint]{aastex}
\usepackage{emulateapj5}
\usepackage{apjfonts}

\begin{document}

\title{
A Failed Gamma-Ray Burst with Dirty Energetic Jets Spirited Away? \\
New Implications for the GRB-SN Connection from Supernova 2002\lowercase{ap}
}

\author{Tomonori Totani\altaffilmark{1, 2}}

\affil{Princeton University Observatory, Peyton Hall, Princeton, NJ 08544-1001, USA }
\altaffiltext{1}{ 
Theory Division, National Astronomical Observatory,
Mitaka, Tokyo 181-8588, Japan }
\altaffiltext{2}{ (Address from June 2003)
Department of Astronomy, Kyoto University, Kitashirakawa, Kyoto 606-8502,
Japan}


\begin{abstract}
The type Ic supernova (SN) 2002ap is an interesting event with very broad
spectral features like the famous energetic SN 1998bw associated with a
gamma-ray burst (GRB) 980425. Here we examine the jet hypothesis from SN
2002ap recently proposed based on the redshifted polarized continuum found in
a spectropolarimetric observation. We show that jets should be moving at
about 0.23c to a direction roughly perpendicular to us, and the degree of
polarization requires a jet kinetic energy of at least $5 \times 10^{50}$
erg, a similar energy scale to the GRB jets. The weak radio emission from SN
2002ap has been used to argue against the jet hypothesis, but we argue that
this is not a problem because the jet is expected to be freely expanding and
unshocked.  However, the jet cannot be kept ionized because of adiabatic
cooling without external photoionization or heating source. We explored
various ionization possibilities, and found that only the radioactivity of
$^{56}$Ni is a plausible source, indicating that the jet is formed and
ejected from central region of the core collapse, not from outer envelope of
the exploding star. Then we point out that, if the jet hypothesis is true,
the jet will eventually sweep up enough interstellar medium and generate
shocks in a few to 10 years, producing strong radio emission that can be
spatially resolved, giving us a clear test for the jet hypothesis.
Discussions are also given on what the jet would imply for the GRB-SN
connection, when it is confirmed.  We suggest existence of two distinct
classes of GRBs from similar core-collapse events but by completely different
mechanisms. Cosmologically distant GRBs having energy scale of $\sim
10^{50-51}$ erg are collimated jets generated by the central activity of core
collapses, associated with $^{56}$Ni ejection along with the jets.  SN 2002ap
can be considered as a failed GRB of this type with large baryon
contamination.  On the other hand, much less energetic ones including GRB
980425 are rather isotropic, which may be produced by hydrodynamical shock
acceleration at the outer envelope.  We propose that the radioactive
ionization for the SN 2002ap jet may give a new explanation also for the
X-ray line features often observed in GRB afterglows.
\end{abstract}


\section{Introduction}
\label{section:intro}
Type Ic supernova (SN) 2002ap has attracted particular attention since its
discovery by Y. Hirose in January 2002, because of its relatively close
distance (about $D = $ 7.3 Mpc, Sharina, Karachentsev, \& Tikhonov 1996; Sohn
\& Davidge 1996) and its broad-line spectral features (Kinugasa et al. 2002)
that are considered as a signature of very energetic supernovae.  Such a
supernova population, often called as hypernovae whose prototype is the
famous type Ic SN 1998bw, have explosion energy more than 10 times larger
than the standard energy ($\sim 10^{51} \rm erg$) when spherical symmetry is
assumed (Iwamoto et al. 1998; Woosley et al. 1999; see also H\"oflich,
Wheeler, \& Wang 1999 for asymmetric modeling of these events with less
extreme explosion energies).  The apparent association of a gamma-ray burst
(GRB) 980425 with SN 1998bw makes these mysterious events even more
interesting in the context of the possible SN/GRB connection.\footnote{GRB
980425 was exceptionally faint one compared with those found at cosmological
distances. However, after the submission of this paper, a supernova SN 2003dh
having similar spectrum to SN 1998bw was discovered to be associated with a
nearby GRB 030329 having normal luminosity (Stanek et al. 2003; Hjorth et
al. 2003), finally confirming the connection between GRBs and energetic
supernovae.}  Mazzali et al. (2002) presented photometric and spectroscopic
modeling of SN 2002ap assuming a spherical explosion, and indicated that the
explosion occurred at Jan 28$\pm$0.5 UT, with a kinetic energy of about
(4--10)$\times 10^{51}$ erg and the progenitor is a C+O star whose main
sequence mass is $\sim 20$--25$M_\odot$. It seems that an interacting binary
is more likely for a star of this mass scale to lose its hydrogen and helium
envelope, but theoretical and metallicity uncertainties do not reject a
single Wolf-Rayet (WR) star as another possible progenitor (Smartt et
al. 2002).

In contrast to SN 1998bw / GRB980425, SN 2002ap was not associated with a GRB
to the sensitivity of IPN (Hurley et al. 2002; but see also Gal-Yam et
al. 2002).  On the other hand, spectropolarimetric observations of SN 2002ap
(Kawabata et al. 2002; Leonard et al. 2002; Wang et al. 2002) give an
interesting hint for hidden energetic ejecta.  Kawabata et al. (2002) noticed
that the spectral shape of polarized continuum observed by Subaru around 10
Feb (i.e., $\sim$13 days after the explosion) apparently looks like the
original unpolarized spectrum, but redshifted by $z = 0.3$ ($\lambda_{\rm
redshifted}/\lambda = 1 + z$) and the ratio of the polarized to unpolarized
flux is $f_P = 0.0018$ (in $f_\lambda$). The polarization angle (PA) is
different from line polarization at this epoch or PA in their second
observation in March (40 days after the explosion).  Interestingly, they got
a consistent PA and wavelength-independent polarization from February to
March, that can be explained simply by an asymmetric photosphere as often seen
in supernova spectra, after they subtracted the redshifted polarized
continuum in February observation.  If it is not a chance coincidence, 
this result can be explained by an asymmetric supernova photosphere {\it and}
a jet moving at a much higher speed ($\sim cz \sim 0.3c$) than the
supernova photosphere (Kawabata et al. 2002). Following this suggestion,
Leonard et al. (2002) confirmed the resemblance between polarized and
redshifted spectrum, by an independent data taken by Keck, although
statistical significance of this resemblance is difficult to assess.

As we will show in \S \ref{section:jet-energy}, jet mass of $ M_{\rm jet}
\sim 0.01 M_{\rm \odot}$ and jet kinetic energy of $ E_{\rm jet} \sim
10^{50-51}$ erg are required to explain the degree of polarization.  It is
quite interesting in the perspective of GRB-SN connection, to note that the
inferred jet energy is very close to the standard energy scale of collimated
jets suggested for GRBs (Frail et al. 2001; Panaitescu \& Kumar 2001). It is
well known that an ultrarelativistic outflow with a Lorentz factor $\gamma
\gtrsim $ 100--1000 is required for successful GRBs to avoid the compactness
problem (Goodman 1986; Paczy\'nski 1986). It may be achieved by production of
a fireball, where enormous energy is injected into a clean region with very
few baryon contamination [$\sim 6 \times 10^{-6} (E/10^{51} {\rm erg})
(\gamma/100)^{-1} M_\odot$]. Then it is naturally expected that there may be
events with similar jet energy but with much larger baryon contamination, and
hence low expansion velocity and no gamma-ray emission, often called as
``failed GRBs'' or ``dirty fireballs'' (e.g., Huang, Dai, \& Lu 2002). 
The energetic jet inferred for SN
2002ap may be the first detection of a failed GRB. On the other hand, even
successful GRBs may also have jets with low velocity and high baryon load,
which are produced in the process of jet acceleration in addition to the
ultra-relativistic component responsible for GRBs. SN 2002ap may have been a
successful GRB, but the jet was not directed to us.

However, this jet hypothesis has been questioned at a few points. Radio
emission is thought as an indicator for the existence of fast moving ejecta,
since it would produce nonthermal synchrotron emission by interaction with
circumstellar matter (CSM) or interstellar matter (ISM).  However, radio
emission of SN 2002ap is much (by more than 3 orders of magnitude) weaker
than that of SN 1998bw, and Berger, Kulkarni, \& Chevalier (2002, hereafter
BKC02) have shown that the observed radio emission can be explained by
synchrotron radiation produced by a spherical ejecta expanding at $\sim 0.3c$
with a total energy of nonthermal electrons of $E_e \sim 1.5 \times 10^{45}$
erg, assuming energy equipartition between electrons and magnetic
fields. These numbers should be compared with those for SN 1998bw, i.e.,
relativistic shock speed with a Lorentz factor $\gamma \sim $ a few and $E_e
\sim 10^{49}$ erg (Kulkarni et al. 1998). 
Based on this result, BKC02 argued that the energetic
jet with $E_{\rm jet} \sim 10^{51}$ erg proposed by Kawabata et al. (2002)
should have produced much stronger radio emission than observed.  Wang et
al. (2002) presented VLT spectropolarimetric observations including earlier
epochs than Keck and Subaru, and found that the continuum polarization
evolved from nearly zero on 3 Feb to 0.2\% on 10 Feb, which is contrary to
what is expected from a simple jet model, since scattering efficiency
should be higher in earlier epochs when jet location is closer to the star. 
Finally, it is uncertain whether
the jet material is kept highly ionized in spite of the expected rapid
adiabatic cooling.

The main purpose of this work is to examine whether the jet hypothesis is
physically tenable especially against the possible difficulties mentioned
above. We found that it is in fact physically possible that the jet exists
but has been spirited away from intensive observational efforts made so far,
except for the Subaru spectropolarimetry.  However, it is still a hypothesis,
and we need a further observational test to prove this. Fortunately, we show
that there is a good test for this hypothesis by future observation; we point
out that a long-term radio monitoring of this object should find re-emergence
of radio emission in a few to 10 years, for which the jet expansion should be
easily resolved by VLBI imaging. Then, we will give discussions on what would
be the implications for the GRB-SN connection if the jet is confirmed in the
future.  Here we also propose a new possible explanation for X-ray line
features often observed in GRB afterglows, which is inspired by the results
of the paper.

The paper is organized as follows. We give an estimate of the jet mass and
energy with a fully relativistic treatment in \S \ref{section:jet-energy}.
We discuss the physical condition of the jet including ionization sources in
\S \ref{section:jet-condition}, and give a detailed modeling of radio
emission from the interaction between the jet and CSM/ISM in \S
\ref{section:radio}. \S \ref{section:discussion} is for discussion on the
GRB-SN connection and X-ray lines in GRB afterglows, and summary and
conclusions are presented in \S \ref{section:conclusion}.

\section{Jet Mass and Energy Estimation}
\label{section:jet-energy}
We assume that the jet is sufficiently collimated and hence
we can define a single scattering angle of photons scattered by the jet,
$\theta_{\rm obs}$, which is the same with
the direction angle of the observer measured from the jet direction.
We also assume that the jet is optically thin for
the electron scattering, which will be checked later. If the jet
is optically thin, the jet mass and energy estimates in this section
do not depend on the jet opening angle.
We use a notation that $x$ and $x'$ are the quantities in
the restframes of the supernova/observer and the jet, 
respectively. [The heliocentric redshift of the host galaxy is 
+631km/s (Smartt \& Meikle 2002) and can be neglected.]
According to the
standard Lorentz transformation, the energy of 
original photons from the supernova has a relation
\begin{eqnarray}
\epsilon'_{\rm org} = \gamma (1-\beta) \ \epsilon_{\rm org} \ , 
\end{eqnarray}
and for the energy of scattered photons by the jet:
\begin{eqnarray}
\epsilon'_{\rm sc} = \gamma (1-\beta \cos \theta_{\rm obs}) \
\epsilon_{\rm sc} \ ,
\label{eq:eps_sc}
\end{eqnarray}
where $\beta$ and $\gamma$ are the bulk velocity and Lorentz factor of
the jet, respectively.
From these relations, and using $\epsilon'_{\rm org} = \epsilon'_{\rm sc}$
for the Thomson scattering, we have the relation between the original
and scattered photon energies:
\begin{eqnarray}
\frac{\epsilon_{\rm sc}}{\epsilon_{\rm org}} = \frac{1}{1+z} = 
\frac{1 - \beta}{1 
- \beta \cos \theta_{\rm obs}} \ ,
\end{eqnarray}
The inverse Lorentz transformation of the equation (\ref{eq:eps_sc}) leads to:
\begin{eqnarray}
\epsilon_{\rm sc} = \gamma (1 + \beta \cos \theta'_{\rm obs}) \
\epsilon'_{\rm sc} \ ,
\end{eqnarray}
and hence $\theta_{\rm obs}$ and $\theta'_{\rm obs}$ are related as:
\begin{eqnarray}
\gamma^2 (1 - \beta \cos \theta_{\rm obs}) (1 + \beta \cos \theta'_{\rm obs})
= 1 \ .
\end{eqnarray}

We should consider three time scales: the time when we observe a
scattered photon ($t_{\rm obs}$, measured from the arrival time of
unscattered photons emitted at the core-collapse date), 
that when the photon is scattered
by the jet ($t_{\rm sc}$), and that when the photon is originally
radiated from the supernova photosphere ($t_{\rm rad}$). The last two
are measured from the core-collapse date in the supernova/observer
restframe. From a geometrical calculation, we find
\begin{eqnarray}
t_{\rm sc} &=& \frac{1}{1 - \beta \cos \theta_{\rm obs}} t_{\rm obs} 
\label{eq:t_sc} \\
t_{\rm rad} &=& \frac{1 - \beta}{1 - \beta \cos \theta_{\rm obs}} t_{\rm obs}
= \frac{1}{1+z} t_{\rm obs}  \label{eq:t_rad}  \ .
\end{eqnarray}

Let $F_{\rm org}(t_{\rm sc}) = L_{\rm org}(t_{\rm rad}) /(4 \pi r_{\rm
jet}^2)$ be the original flux from the supernova at the jet location, $r_{\rm
jet} = c \beta t_{\rm sc}$, and the flux in the jet restframe is
\begin{eqnarray}
F'_{\rm org} = \gamma^2 (1-\beta)^2 F_{\rm org} \ .
\end{eqnarray}
The luminosity of scattered light per unit solid angle is given as:
\begin{eqnarray}
\frac{dL'_{\rm sc}(\theta'_{\rm obs})}{d\Omega} = N_e \frac{d\sigma(
\theta'_{\rm obs})}{d\Omega}  F'_{\rm org} \ ,
\end{eqnarray}
where $N_e = M_{\rm jet} f_{\rm el} / (\mu_e m_p )$ is the free electron
number in the jet, $\mu_e$ the nucleon to electron number ratio, $m_p$ the
proton mass, and $f_{\rm el}$ the fraction of free ionized electrons.
Assuming that the jet material is mostly heavy element, e.g., C+O, we set
$\mu_e = 2$.  Here, $d\sigma(\theta_{\rm obs}')/d\Omega = (3/16\pi) \sigma_T
(1 + \cos^2 \theta_{\rm obs}')$ is the cross section of Thomson scattering
for unpolarized light.  The scattered luminosity $dL'_{\rm sc}/d\Omega$
is related to that in the supernova/observer
restframe as (e.g., Rybicki
\& Lightman 1979):
\begin{eqnarray}
\frac{dL_{\rm sc}(\theta_{\rm obs})}{d\Omega} 
= \frac{1}{\gamma^4 (1 - \beta \cos \theta_{\rm obs})^4} 
\frac{dL'_{\rm sc}(\theta'_{\rm obs})}{d\Omega} \ .
\end{eqnarray}
The ratio of polarized to unpolarized flux, $f_P$, is given as:
\begin{eqnarray}
f_P (\theta_{\rm obs}) &=& \Pi(\theta'_{\rm obs}) \frac{dL_{\rm sc}}{d\Omega}
\left( \frac{L_{\rm org}(t_{\rm obs})}{4 \pi} \right)^{-1} \frac{1}{1+z} \ ,
\end{eqnarray}
where $\Pi = (1 - \cos^2 \theta'_{\rm obs})/ (1 + \cos^2 \theta'_{\rm obs})$
is the degree of polarization of scattered wave, and the last factor of
$(1+z)^{-1}$ is coming from the definition of $f_P$ by the ratio of flux
per unit wavelength, $f_\lambda$.

\begin{figure}
\plotone{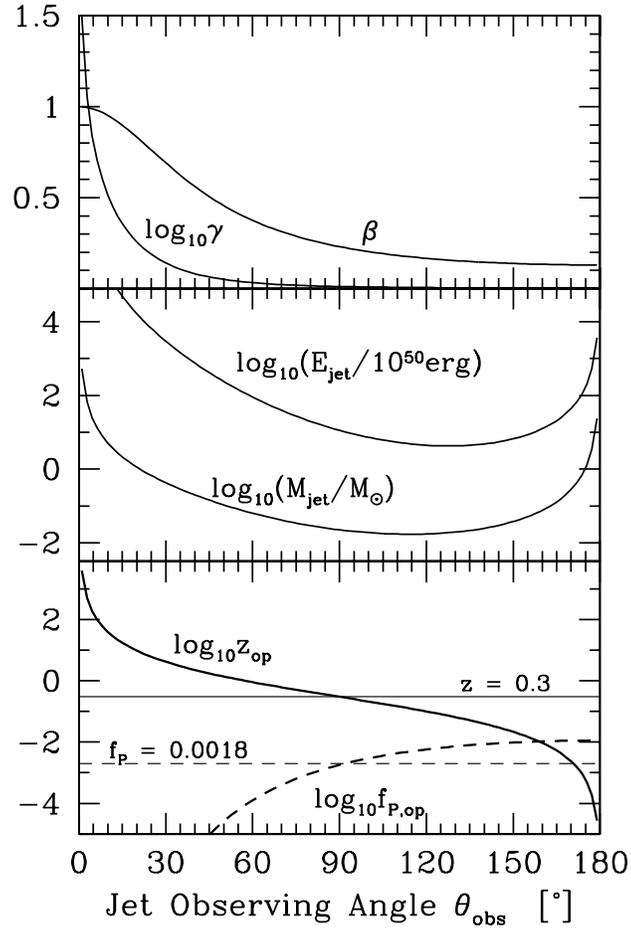}
\caption{Top and middle panels: the jet velocity $\beta$, Lorentz factor
$\gamma$, the jet mass $M_{\rm jet}$, and the jet kinetic energy $E_{\rm
jet}$ required to explain the observed redshift $(z = 0.3)$ and polarization
degree $(f_P = 1.8 \times 10^{-3})$, as a function of the observer's direction
$\theta_{\rm obs}$ measured from the jet direction.  Bottom panel: the
polarization degree $(f_{P, \rm op})$ and redshift ($z_{\rm op}$) expected
for the opposite jet having the direction of $(180^\circ - \theta_{\rm
obs})$, with the same jet parameters shown in the top and middle panels. 
Observed redshift and $f_P$ are marked by horizontal lines.}
\label{fig:jet-energy}
\end{figure}

From eqs. (\ref{eq:t_sc}) and (\ref{eq:t_rad}), 
the fractional time delay of scattered photons from direct 
unscattered photons is $(t_{\rm obs} - t_{\rm rad})/ t_{\rm obs}
= 1 - (1+z)^{-1} = 0.23$, which is small and independent of 
unknown $\theta_{\rm obs}$.
Since the luminosity and spectrum of the supernova are not expected to change
significantly within these time scales, 
we expect that the scattered spectrum is similar to
the unscattered one except for the redshift, as observed. Therefore
we do not have to take into account the luminosity and spectral evolution
of supernova, i.e., $t_{\rm rad} \sim t_{\rm obs}$.
In the top and middle panels of Fig. \ref{fig:jet-energy}, we show
$\beta$, $\gamma$, $M_{\rm jet}$, and $E_{\rm jet} = M_{\rm jet} c^2 (\gamma -
1)$ required to reproduce the observed redshift ($z = 0.3$) and degree of
polarization ($f_P = 1.8 \times 10^{-3}$), as a function of the jet
viewing angle,
$\theta_{\rm obs}$. Here, we have assumed $f_{\rm el} = 0.3$ and
$t_{\rm obs} = 10$d, and scaling of the results by 
different values of these parameters is obvious.

It is likely that there is another jet in the opposite
direction from the supernova, in addition to the jet considered so far, and
the opposite jet should also produce another redshifted polarization
component. Since the data show only one redshifted component, the
contribution from this opposite jet must be with similar polarization degree
and redshift to the original jet, or negligibly small due to too small $f_P$
or very large redshift.  In the bottom panel of 
Fig. \ref{fig:jet-energy}, we plot the redshift ($z_{\rm op}$) and the
polarization degree ($f_{P, \rm op}$) by the opposite jet having the same
jet velocity and mass but $\theta_{\rm obs, op} = 180^\circ - \theta_{\rm
obs}$.  This gives a constraint of $\theta_{\rm obs} \lesssim 100^\circ$,
otherwise we should have observed another polarized continuum component
with larger $f_P$ and smaller redshift than the observed ones.

The jet velocity is roughly constant at $\beta \sim 0.2$ for $\theta_{\rm
obs} \gtrsim 90^\circ$, but it becomes more relativistic with decreasing
$\theta_{\rm obs}$ at $\theta_{\rm obs} \lesssim 90^\circ$, because the
redshift effect of the jet motion is compensated by the blueshift to the
observer.  The jet mass and energy rapidly becomes larger with decreasing
$\theta_{\rm obs}$, mainly because of less efficient polarization and larger
$r_{\rm jet} \propto t_{\rm sc} \gg t_{\rm obs}$.  Small $\theta_{\rm obs}
(\lesssim 60^\circ)$ seems not favored from energetics, since it requires jet
energy of more than $10^{52}$ erg.  These considerations lead to a conclusion
that the jet directions must be close to $\theta_{\rm obs} \sim 90^\circ$
(probably within $\pm$ 10--20$^\circ$) and both the two jets contributed
roughly equally to the observed polarization.  Therefore we assume two jets
with $\theta_{\rm obs} = 90^\circ$ in this work. Then we found $\beta = 1 -
(1+z)^{-1} = 0.23$, and the jet mass is
\begin{eqnarray}
M_{\rm jet} = 0.011 \left(\frac{f_P}{0.0018}\right)  
\left(\frac{t_{\rm obs}}{10 \rm d}\right)^2 
\left( \frac{f_{\rm el}}{0.3}\right)^{-1} M_\odot \ ,
\end{eqnarray}
where $M_{\rm jet}$ is re-defined as the mass of each jet.  Therefore,
observed redshifted polarization can be explained if the jet material is
modestly ionized, with the kinetic jet energy $E_{\rm jet} \sim 5 \times
10^{50} (M_{\rm jet}/0.01 M_\odot) (\beta /0.23)^2$ erg. In the following of
this paper we consistently use $\theta_{\rm obs} = 90^\circ$, $t_{\rm sc} =
t_{\rm obs}$, $\beta = 0.23$, and $M_{\rm jet} = 0.01 M_\odot$.

\section{Physical Conditions of the Jet and Ionization}
\label{section:jet-condition}
\subsection{Jet is freely expanding}
First we consider the fact that the radio emission from SN 2002ap was very
weak. If a considerable part of the jet kinetic energy was converted into
nonthermal electrons via shock acceleration, inevitably there must be very
strong radio emission which should have been even stronger than SN
1998bw. However, if the amount of CSM swept-up the jet is much smaller than
the jet mass, the jet feels almost no deceleration and the majority of the
jet material remains unshocked. We do not expect radio emission from such an
almost freely expanding jet, and we only expect radio emission by CSM
swept-up by the jet.  This emission and total energy of radio emitting
electrons ($\sim 10^{45}$ erg) are simply related by the jet velocity and CSM
density, and {\it not} related with the total jet mass and kinetic
energy. Stellar wind mass-loss rate of Wolf-Rayet stars, which are considered
as a possible candidate of the SN Ic progenitors, is typically $\dot{M}_w
\sim 10^{-6}$--$10^{-5} M_\odot$/yr with a wind velocity of $V_w \sim 10^3$
km/s (McCray 1983; Garc\'ia-Segura, Mac Low, \& Langer 1996; Garc\'ia-Segura,
Langer, \& Mac Low 1996).  It is generally assumed that the CSM around radio
supernovae has a stellar wind profile, i.e., $\rho_{\rm CSM}(r) = \dot{M}_w /
(4 \pi r^2 V_w)$.  Then the swept-up mass by the jet becomes $M_{\rm sw} = b
\dot{M}_w r_{\rm jet} / V_w = 1.9 \times 10^{-7} b_{-1} \dot{M}_{w, -6} V_{w,
3}^{-1} t_{10} M_\odot$, where $b = 0.1 b_{-1} = \Omega_{\rm jet}/4\pi$ is
the beaming factor of the jet opening angle, $t_{10} = t_{\rm obs}$/10d,
$\dot{M}_{w, -6} = \dot{M}_w / (10^{-6} M_\odot \rm /yr)$, and $V_{w, 3} =
V_w/(10^3 \rm km/s)$.  This is much smaller than the jet mass and hence the
jet is not decelerated.

Radio emission from SN 1998bw and 2002ap can be explained by isotropic high
speed ejecta interacting with the CSM density consistent with the wind
parameters similar to the above values (Kulkarni et al. 1998; 
Li \& Chevalier 1999; BKC02).  [The
speed of ejecta is, however, 
considerably different for these two; $\gamma_{\rm ej} \sim$ a few for the
former but $\upsilon_{\rm ej} \sim 0.3c$ for the latter.]  Since the inferred
jet velocity of SN 2002ap is close to $\upsilon_{\rm ej}$, and the swept-up
CSM mass by the jet should be smaller than that by isotropic ejecta
because of the collimation, it looks
reasonable that the radio emission from the external shock front of the jet is
equal to or smaller than the observed radio emission. We will present more
detailed radio emission modeling in \S \ref{section:radio}.

Therefore the weak radio flux from SN 2002ap does not immediately exclude the
jet hypothesis, if it is expanding freely. It may also be useful to recall
that the kinetic energy of supernovae ($\sim 10^{51}$ erg) is hardly
converted into radiation, but supernovae are heated and shining by
radioactivity.  However, free expansion raises another problem because of
the expected rapid adiabatic cooling. The jet material must be ionized at
least modestly, but as we will show below, the adiabatic cooling is so strong
that the temperature of the jet material is likely lower than that necessary
to keep it ionized. Therefore we need an external radiation or heating source
of ionization. We will discuss this issue in detail below.

\subsection{Optical depth and thinning burst}
First we check the optical depth of the jet material to the 
photon-electron scattering. The jet must be mostly transparent
to radiation at the day $\sim 13$, 
for the jet material to contribute the scattering 
and polarization efficiently. It can be written as:
\begin{eqnarray}
\tau_{\rm jet} \sim \frac{f_{\rm el} M_{\rm jet} }{\mu_e m_p} \frac{\sigma_T }{
4 \pi b r_{\rm jet}^2}
= 0.088 \ f_{\rm el} \ b_{-1}^{-1} \ t_{10}^{-2} \ .
\end{eqnarray}
Therefore
the jet becomes optically thin at a time $t_{\rm th} \sim 3.0 \
f_{\rm el}^{1/2} \ b_{-1}^{-1/2} $ days. This thinning time is
close to the epoch when the continuum polarization evolved from nearly
zero to the 0.2\% level in the VLT observation, and hence indicating that
the initially unobserved polarization can be explained by high optical
depth of the jet. 

In analogy to the fireball theory for GRBs, we expect a burst of
radiation when the jet becomes optically thin (M\'esz\'aros, Laguna,
\& Rees 1993), and here we examine how much radiation we expect from this.
Suppose that the jet is in thermal equillibrium and the total internal
energy is comparable with the kinetic energy of the jet, $E_{\rm jet}$,
at the initial radius $r_i$. If the jet interacts with stellar envelope
and dissipate its kinetic energy, the radius of the progenitor C+O
star, $r_i \sim 10^{10}$ cm is a reasonable choice, while another possibility
is $r_i \sim 10^6$ cm if the jet is directly emitted from the central 
compact neutron star or black hole. The initial
temperature $T_i$ is determined by the internal energy density $U_i = 
E_{\rm jet}/V_i$, and both the internal energy and temperature 
adiabatically decrease as $\propto (r_{\rm jet}/r_i)^{-1}$, 
where $V_i = 4 \pi \zeta r_i^3$ is the initial volume
of the jet material and $\zeta$ is the fractional thickness of the jet shell.
Here we assumed that the jet is homologously expanding like supernova ejecta,
and hence $V \propto r_{\rm jet}^{3}$. Then we find the internal
energy and temperature at the thinning time $t_{\rm th}$ as:
\begin{eqnarray}
E_{\rm th} &\sim& 2.8 \times 10^{45} f_{\rm el}^{-1/2} b_{-1}^{1/2} 
r_{\rm i, 10} \ \rm erg \\
k_B T_{\rm th} &\sim& 0.41  f_{\rm el}^{-1/2} b_{-1}^{1/4} \zeta_{-1}^{-1/4}
r_{\rm i, 10}^{1/4} \ \rm eV  \ ,
\end{eqnarray}
where $\zeta_{-1} = \zeta / 0.1$ and $r_{i, 10} = r_i/(10^{10}$cm).
The flux of this thinning radiation is $F_{\rm th} \sim E_{\rm th}
/ (4 \pi D^2 t_{\rm th}) = 1.7 \times 10^{-12} f_{\rm el}^{-1} 
b_{-1} r_{\rm i, 10}
\ \rm erg \ cm^{-2} s^{-1}$, which is sufficiently smaller than the
observed UV--optical--IR flux of $\sim 2 \times 10^{-10} \ \rm erg
\ cm^{-2} s^{-1}$ at the day 5 (Sutaria et al. 2003).

If there is no external heating or ionizing radiation source,
the ionization balance of the jet would be determined by 
collisional ionization coefficient $q_{\rm col}$ by thermal electrons 
and radiative recombination coefficient $\alpha_{\rm rec}$.
Comparing coefficients of these processes (Lotz 1967; Nahar \&
Pradhan 1997; Nahar 1999),
the temperature must be higher than $\sim 5$ eV for the
C+O matter to be ionized doubly or more. Considering the temperature
derived above, and the weak dependence on the unknown parameters,
it seems unlikely that the jet is sufficiently hot to keep itself
ionized. After the jet becomes optically thin, radiative cooling may
further decrease the temperature.
Then we need an external source of heating or ionizing photons.

\subsection{Photoionization}

To begin with, we estimate the total ionizing photon luminosity required to 
keep all the jet material ionized (a similar argument used to derive the
Str\"omgren radius). This is given by the recombination rate as: 
\begin{eqnarray}
L_{\rm ph, rec} &\sim& \alpha_{\rm rec} n_{\rm e} N_{\rm ion} 
\nonumber \\ &=& 
5.6 \times 10^{50} b_{-1}^{-1} \zeta_{-1}^{-1} \mu_{14}^{-1} t_{10}^{-3} 
\nonumber \\
&\times&
\left(\frac{\alpha_{\rm rec}}{10^{-11} \ \rm cm^3 s^{-1}}\right)^{-1}
\left(\frac{f_{\rm el}}{0.3}\right) \ \rm s^{-1} \ ,
\label{eq:L_rec}	
\end{eqnarray}
where $n_e$ is the free electron number density, $N_{\rm ion}$
the number of ions, and $\mu_{\rm ion} = 14 \mu_{14}$ is
the mean molecular weight of ions.
Correspondingly, the ionizing flux in the jet must be stronger than
$F_{\rm ion} \sim L_{\rm ph, rec} / (4 \pi b r_{\rm jet}^2)$, 
and the abundance ratio
between species $X^{+i+1} + e^{-} \leftrightarrow X^{i}$ is given as:
\begin{eqnarray}
\frac{N(X^{+i+1})}{N(X^i)} &=& \frac{\sigma_{\rm ion} F_{\rm ion}}{
\alpha_{\rm rec} n_e} \\
&=& 9.4 \times 10^4 b_{-1}^{-1} \mu_{14}^{-1} t_{10}^{-2}
\left(\frac{\sigma_{\rm ion}}{5 \times 10^{-18} \ \rm cm^2}\right) \ .
\end{eqnarray}
Here, we used typical values of $\alpha_{\rm rec}$ and 
ionization cross section $\sigma_{\rm ion}$
for C$^+$ $\leftrightarrow$ C$^{+2}$ or O$^+$ $\leftrightarrow$ O$^{+2}$
at temperature $\sim$ 1 eV (Osterbrock 1989).  
Therefore, a necessary and sufficient condition for ionization is that
the jet matter is radiated with a total photon luminosity given in eq.
(\ref{eq:L_rec}).

First we consider the possibility of ionization by radiation from the
supernova.  The effective temperature inferred from optical colors of SN
2002ap at day $\sim 10$ is about $T_{\rm eff} \sim $ 6,000 K (Gal-Yam et
al. 2002), and the number of photons in the black-body tail above $\nu_T \sim
6 \times 10^{15}$ GHz, which is a threshold frequency of ionizing photons for
CII and OII, is quite small.

There is an XMM-Newton observation of SN 2002ap on Feb 2 UT, i.e., about 5
days after the explosion (Soria \& Kong 2002; Sutaria et al. 2003). The flux
in 0.3--10 keV is $1.07 \times 10^{-14} \rm erg \ cm^{-2} s^{-1}$, and the
spectrum can be fit with a power-law with a photon index of $\alpha_X =
2.6^{+0.6}_{-0.5}$ $(f_\nu \propto \nu^{-\alpha_X+1})$. The flux extrapolated
down to UV bands may ionize the jet, and softer spectrum gives stronger
ionization flux. 

However, following arguments give further constraints on the possible
range of $\alpha_X$.
Sutaria et al. (2003) ascribed this X-ray flux to be inverse-Compton
scattering of the optical photons, and in this case the extrapolated
nonthermal flux down to the optical bands should be about $\tau_{\rm CSM}
F_{\rm opt}$, where $\tau_{\rm CSM}$ is optical depth of hot electrons in
shocked CSM around the supernova and $F_{\rm opt}$ the optical flux of the
supernova (Fransson 1982; Chevalier \& Fransson 2001; Sutaria et al. 2003).
The hot CSM gas responsible for X-ray emission is swept up either by the jet
or by another isotropic fast ejecta as considered by BKC02 (see also \S
\ref{section:radio} for examination of these two possibilities from radio
data).  In either case, the velocity of the shock front is 0.23--0.3$c$, and
hence the location of X-ray emitting region is $r_X \sim 6.0 \times 10^{15}
t_{10}$ cm.  Therefore we find $\tau_{\rm CSM} = 2.0 \times 10^{-5}
\dot{M}_{w, -6} V_{w, 3}^{-1} t_{10}^{-1} b_X$, where $b_X$ is the sky
coverage of X-ray emitting region viewed from the supernova photosphere; $b_X
= 1$ for the isotropic ejecta while $b_X = b$ for the jet.  As we will show
in \S \ref{section:radio}, $\dot{M}_w \lesssim 10^{-4} M_\odot$/yr is
required for successful modeling of the observed radio data. Then, comparing
with the observed optical flux, $\alpha_X \lesssim 2.6$ and 2.2 is required
for $b_X = 1$ and $b_X = b = 0.1$, respectively, even though softer index is
allowed by the observational error.  BKC02, on the other hand, suggested that
the X-ray flux is explained by the same synchrotron radiation as the radio
observations. The radio and X-ray fluxes at the day 5 are connected by a
power law with a photon index of $\alpha_X' = 1.5$, and considering that the
photon index changes by 0.5 below and above the cooling break frequency, the
maximum photon index allowed in the X-ray band is $\alpha_X < \alpha_X' + 0.5
= 2$, irrespective of $b_X$.

Now we compare these constraints on $\alpha_X$ with the range required to
ionize the jet.
The observed X-ray flux at the day 5 should not be much different at the day
$\sim$10--13, and we extrapolate the X-ray luminosity down to the UV band and
compare to $L_{\rm ph, rec}$. Note that only a fraction of the X-ray
luminosity is directed to the jet material, and this fraction is given by
$\sim b/b_X$ from a geometrical consideration. We found that the spectral
index must be extremely soft as $\alpha_X \gtrsim$ 5 or 4 for $b_X = $ 1 or
$b_X = b = 0.1$, respectively, in order that the extrapolated flux down to
$\nu_T$ is equal to $L_{\rm ph, rec}$. Therefore we can safely exclude the
possibility that the nonthermal radiation producing the observed X-rays is
ionizing the jet.

Secondly we consider a possibility that a hot, UV-radiating star close to the
SN 2002ap may ionize the jet. It is expected that SN 2002ap occurred in a
massive star forming region where young massive stars are clustering. A close
binary system is a candidate for the type Ic supernova progenitors
(Nomoto, Filippenko, \& Shigeyama 1990), and it
may provide even stronger ionization source. However, the total ionization
luminosity given in eq. (\ref{eq:L_rec}) is even larger by a factor of several
than the ionization flux, $\sim 10^{49.5} \ \rm s^{-1}$, above the frequency
$\nu_T \sim 6 \times 10^{15}$ Hz for the most luminous and hottest stars
(Schaere \& de Koter 1997). It should be noted that this luminosity is for
all direction, but only the radiation within the solid angle of the jet
viewed from the ionizing star is available for the jet ionization, which is
expected to be a small fraction.  If the region around SN 2002ap is filled up
by radiation field with $\sim F_{\rm ion}$ to a radius of $r_{\rm jet}$, the
region should have a luminosity of at least $4 \pi F_{\rm ion} r_{\rm
jet}^2$, corresponding bolometric luminosity of $2.1 \times 10^8 b_{-1}^{-2}
\zeta_{-1}^{-1} \mu_{14}^{-1} t_{10}^{-3} (f_{\rm el}/0.3) L_\odot$, assuming
the spectral energy distribution of the most luminous O stars. Such a huge
luminosity is apparently ruled out by the prediscovery image of the SN 2002ap
field reported by Smartt et al. (2002). To conclude, ionization by nearby
young stars is impossible.

\subsection{Radioactive heating and collisional ionization}
\label{section:nickel}
Since photoionization of the jet seems difficult, the only way to ionize the
jet is enhanced collisional ionization by external heating.  If the jet is
generated at the central compact object, it might include a significant
amount of radioactive nuclei such as $^{56}$Ni. Asymmetric explosion induced
by the jet should also affect nucleosynthesis, and $^{56}$Ni production
along the jet direction is enhanced (Nagataki 2000; Maeda et
al. 2002).  $^{56}$Ni decays by electron capture and gamma-ray emission to
$^{56}$Co, with an exponential decay time scale of $t_{\rm Ni} = t_{1/2} /
\ln (2) = 8.5$ d and decay energy is $\epsilon_{\rm Ni} = 2.1$ MeV. When the
material is optically thick, the radioactive heat is quickly thermalized into
optical radiation field, as generally seen for supernovae. On the other hand,
if the material is mildly optically thin, the gamma-rays emitted by decaying
$^{56}$Ni scatter electrons with a probability $\sim \tau_{\rm jet}$, and
since the gamma-ray energy is comparable with the electron rest mass, the
scattered electrons acquire mildly relativistic speed and energy.  Such high
energy electrons would lose their energy by ionization loss in the jet
plasma, with a time scale of
\begin{eqnarray}
t_{\rm il} 
&\sim& 2.0 \times 10^4 (\upsilon_e/c)^4 
t_{10}^3 \zeta_{-1} b_{-1} \ \rm s, 
\end{eqnarray}
where $\upsilon_e$ is the initial velocity of high energy electrons.
Here we used the ionization loss formulae of
Longair (1992) and the logarithmic factor is set to be 15. Therefore the
energy deposited by radioactive gamma-rays
is used to ionize the jet material within the time scale
of interest, giving an efficient ionization process.
When optical depth is very low, this process would be
dominated by positrons emitted from decay of $^{56}$Co, which has a longer
exponential lifetime of $t_{\rm Co} = $ 111.26 days and energy fraction given
to positrons is 3.5\% of the total decay energy (Arnett 1979; Woosley, Pinto,
\& Hartmann 1989).

The ionizing
balance is determined by the energy balance between radioactive heating
and recombination cooling (see also Graham 1988) as:
\begin{equation}
\tau_{\rm jet} 
\frac{ f_{\rm Ni} M_{\rm jet} }{\mu_{\rm Ni} m_p } \  \epsilon_{\rm Ni}
\frac{\exp(-t_{\rm sc}/t_{\rm Ni})}{t_{\rm Ni}}  \gtrsim \alpha_{\rm
rec} n_e N_{\rm ion} w \ ,
\end{equation}
where $f_{\rm Ni}$ is the $^{56}$Ni mass fraction in the jet, and
$w$ the ionization potential.
The recombination rate coefficient depends on the electron gas temperature,
which is determined by balance between the radioactive heating and cooling
processes. We can estimate the minimum amount of $^{56}$Ni by
taking the minimum value of $\alpha_{\rm rec}$ as
\begin{equation}
f_{\rm Ni} \gtrsim 0.68 \ t_{10}^{-1} \zeta_{-1}^{-1} e^{(t-10\rm d)/t_{\rm
Ni}} \mu_{14}^{-1} w_{20}
\left( \frac{\alpha_{\rm rec}}{3 \times 10^{-12}
\rm \ cm^3 s^{-1}} \right)
\end{equation}
where the adopted value of $\alpha_{\rm rec}$ is minimum value of doubly
ionized oxygen or carbon at temperature of $\sim 10^4$K (Nahar \& Pradhan
1997; Nahar 1999) and $w_{20} = w$/(20 eV).  The jet may include significant
amount of heavier nuclei that are difficult to ionize for a fixed value of
$f_{\rm el}$, but on the other hand, it may also include considerable helium
that is easier to ionize. The helium could be mixed from remaining helium
layer of the progenitor, or it may be newly synthesized. Production of helium
is also enhanced along the jet direction in energetic jet-like
nucleosynthesis (Maeda et al. 2002). We also note that highly ionized heavy
nuclei, such as $^{56}$Ni, should produce observable line emission in X-ray
bands, and hence the observed weak X-ray flux gives a constraint on the
species of ionized elements. (See \S \ref{section:X-ray-line} for possible
connection to X-ray line features often observed in GRB afterglows.)
Whatever the jet composition is, the above result indicates that, if the jet
is kept ionized by radioactive heating, it must have a considerable amount of
$^{56}$Ni (mass fraction of order unity).  This estimate is, however, very
uncertain especially about the composition of the jet, $\alpha_{\rm rec}$ and
electron temperature.  More sophisticated treatment is necessary to determine
the ionization status, but it is beyond the scope of the paper.  Therefore,
it is difficult to conclude that the jet should be ionized, but it seems the
best candidate of ionization process among others.

The jet may be ionized by gamma-rays of $^{56}$Ni decay leaking from the
photosphere of SN 2002ap, even if the jet does not have radioactive
nuclei. In fact, ionization of helium envelope above the photosphere, which
is required to explain the observed He lines in SN 1987A and type Ib
supernovae, is ascribed to the leaking gamma-rays from photosphere (Graham
1988; Lucy 1991). The mass of $^{56}$Ni produced by SN 2002ap is estimated to
be $0.07 \pm 0.02 M_\odot$ from the light curve modeling by Mazzali et
al. (2002), which is larger than the jet mass.  However, the efficiency for
gamma-rays to hit the jet is reduced by the beaming factor, $b$, and it is
further reduced by escaping fraction from photosphere. Although some
supernovae, including SN 1998bw, showed evidence that a significant amount of
gamma-rays are leaking in late phase ($\gtrsim$ 30 days) (Nakamura et
al. 2001; Patat et al. 2001), the leaking fraction should not be large in
early phase of $\sim$ 10 days, when the optical luminosity is still glowing
up by diffusion of radioactive heat to the photosphere. According to the
model of Mazzali et al. (2002), about 10\% of gamma-rays are leaking, most of
which are produced by $^{56}$Ni outside the photosphere, at the day $\sim 10$
(K. Maeda \& K. Nomoto 2003, a private communication).  This should be
considered as an upper limit for the leaking fraction since the model assumes
the maximally possible mixing, i.e., uniform distribution of $^{56}$Ni. [The
best-fit model of Mazzali et al. (2002) has less significant mixing.]

Therefore, leaking gamma-rays from or around the photosphere seem
less efficient than $^{56}$Ni in the jet, if the $^{56}$Ni distribution
is what is expected from isotropic modeling. However,
if the explosion is very asymmetric due to the jet formation activity
and considerable amount of $^{56}$Ni is ejected outside the photosphere,
it may ionize the jet. It should also
be noted that SN 1998bw produced much larger amount of $^{56}$Ni
($\sim 0.7 M_\odot$, Iwamoto et al. 1998) than SN 2002ap.
We cannot reject that a comparable $^{56}$Ni was produced also in SN 2002ap,
but most of it is well outside the 
photosphere where optical depth is low and radioactive decay energy 
mostly escapes as gamma-rays, not in optical bands. Such $^{56}$Ni could be
missed in the modeling by Mazzali et al. (2002) based on optical
observations. Such extreme mixing and distribution of $^{56}$Ni
is unlikely to occur simply by hydrodynamical instability in C+O stars
(K. Maeda \& K. Nomoto 2003, a private communication, see also Shigeyama et
al. 1990). Hence, significant ejection of $^{56}$Ni from the stellar core
by jet formation activity is again indicated.

\section{Radio Emission by Swept-up Material}
\label{section:radio}
\subsection{Early emission by interaction with presupernova wind}
The observed radio emission is considered to be produced by shocked CSM
swept-up by high velocity supernova ejecta.  Then there are two
possibilities: (i) the observed radio flux is generated by CSM swept up by
the jet responsible for the redshifted polarization, or (ii) the radio flux
is from CSM swept up by isotropic supernova ejecta that is a different
component from the jet, as considered by BKC02, and the radio emission by CSM
swept up by the jet was weaker than observed.  The former option predicts
that the shock front of the radio emission region is not decelerating,
otherwise the shock generated in the jet material would overproduce much the
observed radio flux.  On the other hand, since the expansion velocity
of isotropic shell considered by BKC02 is similar to that of the jet,
the mass of the isotropic ejecta in the
latter option must be much smaller than the jet mass, otherwise the
isotropic component would spoil the redshifted polarization produced by
scattering in the jet. The mass of isotropic component can be as small as the
swept-up CSM, that is much smaller than the jet mass as argued above, and it
should be decelerated by swept-up CSM. 
BKC02 found that the radio data is not sufficient to
constrain whether the shock radius is decelerating or not, and hence cannot
constrain this possibility.  Here we examine the possibility (i) by a
detailed modeling of the observed radio emission with a collimated jet.

Given the jet opening angle and the density of CSM (determined by $\dot{M}_w$
and $V_w$), we can calculate the mass and energy density of shocked CSM swept
up by the jet moving at a constant velocity, $0.23c$. We assumed the strong
shock limit with a compression factor of 4, and temperature of the shocked
CSM is calculated by the standard shock theory for supersonic piston.  Then
we can calculate the synchrotron flux according to the standard formulae, if
fractional energy densities of nonthermal electrons ($\epsilon_e$) and
magnetic field ($\epsilon_B$), electron power index ($p, dN_e/d\gamma_e
\propto \gamma_e^{-p}$), and the minimum Lorentz factor of nonthermal
electrons ($\gamma_m$) are specified. We calculated the synchrotron flux
taking into account synchrotron self-absorption (SSA) by formulations given
in Li \& Chevalier (1999), and also free-free absorption (FFA) by
formulations in Weiler et al. (1986), assuming pre-shocked CSM temperature
$T_{\rm CSM} = 10^4$K. We fix $\epsilon_e = 0.05$, and find best-fit
parameters of $\dot{M}_w$ and $\epsilon_B$ to the observed radio data (BKC02)
by $\chi^2$ analysis, as a function of the beaming factor $b$. The radio flux
is showing an evidence of modulation, presumably due to the interstellar
scattering and scintillation (ISS), and the minimum $\chi^2$ is unacceptably
large without this effect taken into account.  Here we calculate ISS
modulation index with parameters given in BKC02, and it is added to the
observational flux errors as quadratic sum.  The wind velocity is fixed to
$V_w = 10^3$ km/s, and different values of $V_w$ simply rescale $\dot{M}_w$
via the CSM density ($\propto \dot{M}_w / V_w$). The change of $\epsilon_e$
is also mostly canceled by scaling of $\dot{M}_w$, except for the strength of
the free-free absorption. We checked that changing $\epsilon_e$ by one order
of magnitude does not affect conclusions derived below. The magnetic field
strength can be expressed as: $B = 0.21 \ \dot{M}_{w, -6}^{1/2} V_{w,
3}^{-1/2} (\epsilon_B/0.01)^{1/2} \ \rm G$.

\begin{figure}
\plotone{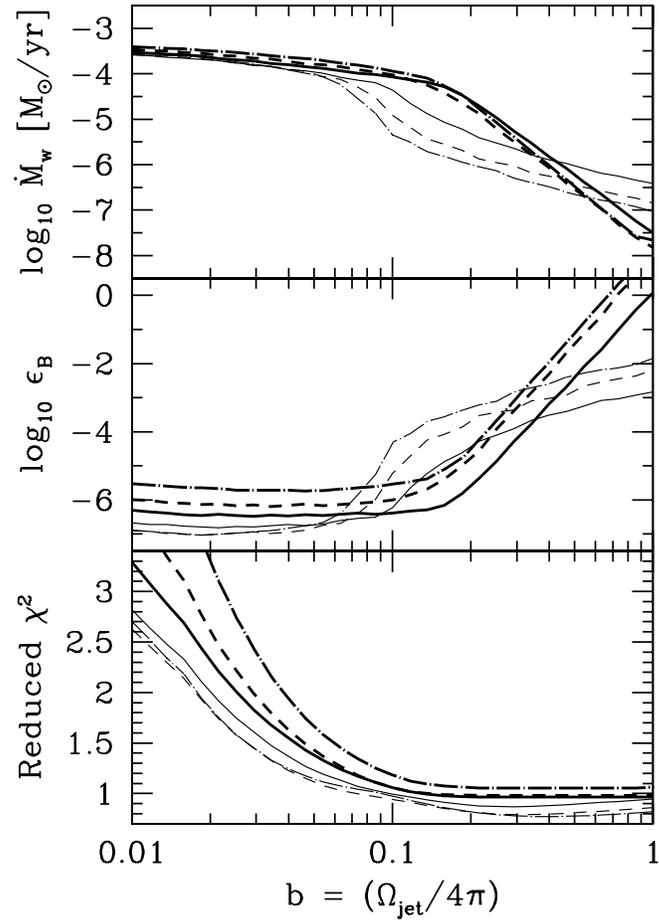}
\caption{
Presupernova mass loss rate $\dot{M}_w$ (top panel),
fractional energy
density of magnetic field (middle panel),
and reduced $\chi^2$ obtained by $\chi^2$ fitting 
of the radio emission expected from swept-up CSM by the jet
to the observed data. The solid, dashed, and dot-dashed curves
are for different electron power index as $p = $ 2.2, 2.5, and 2.8,
respectively. The low and high values of the minimum electron 
Lorentz factor, ($\gamma_{m, l}$ and $\gamma_{m, h}$), are used
for thick and thin curves, respectively.
}
\label{fig:eps_B_Mwind}
\end{figure}

\begin{figure}
\plotone{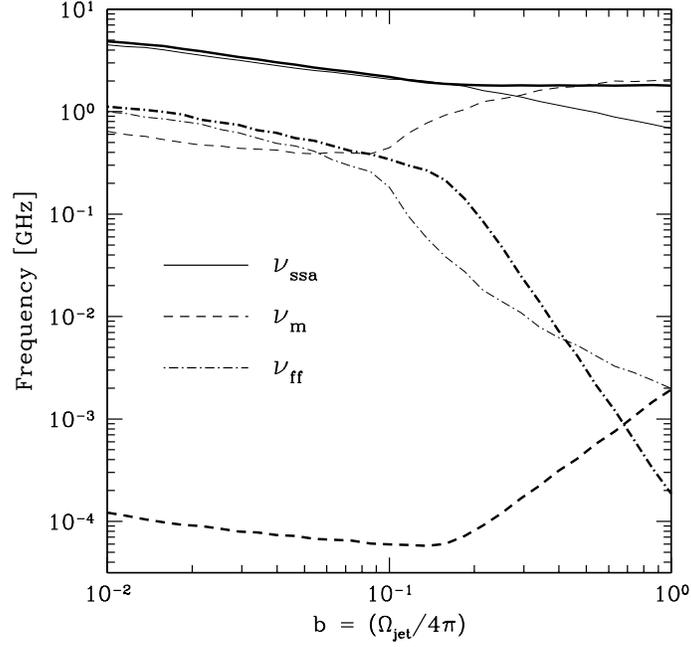}
\caption{The characteristic synchrotron frequency $\nu_m$ (dashed),
synchrotron self absorption frequency
$\nu_{\rm ssa}$ (solid), and free-free absorption frequency $\nu_{\rm ffa}$ 
(dot-dashed) at the
day 7, for the best-fit models to the observed radio flux and spectrum of SN
2002ap. The electron power index is fixed to $p = 2.2$, and thick and thin
curves are for the low and high values of $\gamma_m$: $\gamma_{m, l}$
and $\gamma_{m, h}$, respectively.  }
\label{fig:nu}
\end{figure}

\begin{figure}
\plotone{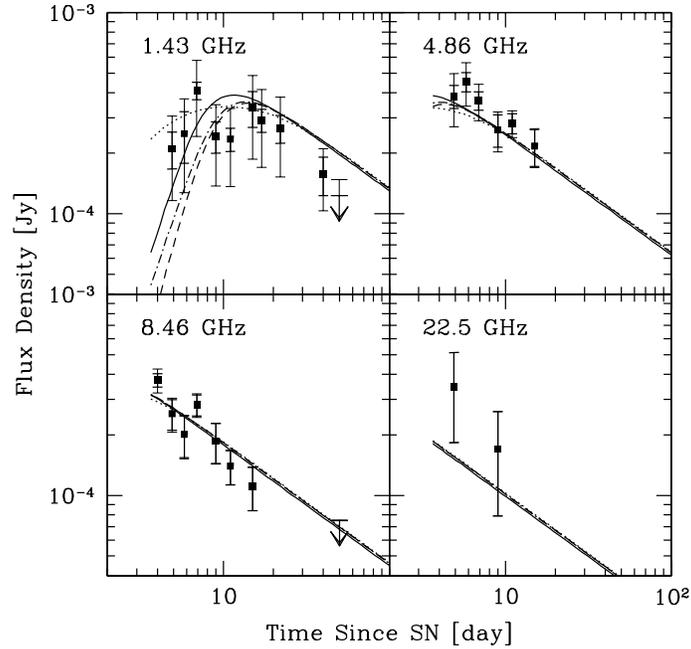}
\caption{The radio flux evolution of the best-fit models for 
$(b, \gamma_m) = (1, \gamma_{m, l})$,
$(1, \gamma_{m, h})$, 
$(0.1, \gamma_{m, l})$, and
$(0.1, \gamma_{m, h})$, for the solid, dotted, dashed, and dot-dashed lines,
respectively. The observed data are from BKC02, with the thick error bars
showing observational errors while the thin error bars include
the ISS modulation by quadratic sum to the observational errors.
}
\label{fig:flux_early}
\end{figure}

The best-fit $\dot{M}_w$ and $\epsilon_B$, as well as the $\chi^2$ value, are
given in Fig. \ref{fig:eps_B_Mwind}.  Here we used two extreme values of
$\gamma_m$: a low value $\gamma_{m, l} = \gamma$ and a high value $\gamma_{m,
h} = 1 + \mu_e m_p (\gamma - 1) / m_e$ for the thick and thin lines,
respectively.  The former corresponds to a case that the electron minimum
energy simply reflects the velocity of the shock, while the latter to a case
that the kinetic energy of ions is efficiently transfered to electrons. We
also used three values of $p = 2.2, 2.5$, and 2.8. The characteristic
synchrotron frequency ($\nu_m$) corresponding to $\gamma_m$, the SSA
frequency ($\nu_{\rm ssa}$), and the FFA frequency ($\nu_{\rm ffa}$) at the
day 7 in these results are given in Fig. \ref{fig:nu}, but only for the $p =
2.2$ case.  The $\chi^2$ degree of freedom is $n_{\rm dof} = 24 - 3 = 21$,
and the minimum reduced $\tilde{\chi}^2 \equiv \chi^2/n_{\rm dof} $ is less
than the unity, i.e., an acceptable fit. The confidence limit projected on
the parameter $b$ can be estimated by a region where $\Delta \chi^2$, i.e.,
difference of $\chi^2$ from the minimum, is smaller than a certain value;
$\Delta \tilde{\chi}^2 < 0.19$ and 0.32 for 95.4 and 99\% C.L., respectively,
assuming a pure Gaussian statistics (e.g., Press et al. 1992). Therefore, we
conclude that a mild beaming $b \sim 0.1$ is marginally allowed and stronger
beaming is excluded for the possibility (i). The flux evolution and comparison
with observed data are shown in Fig. \ref{fig:flux_early}, for the best-fit
models with $b = 0.1$ and 1. The result in the isotropic case ($b = 1$)
is similar to that of BKC02, as it should be.

A general trend seen in Fig. \ref{fig:eps_B_Mwind} can be understood as
follows. When jet is more strongly collimated, the amount of CSM swept-up by
the jet becomes smaller, and hence higher mass loss rate is required to
compensate this. However, the observed spectral feature is mostly explained
by SSA, and hence magnetic field must become smaller to keep SSA frequency at
the observed value.  This explains behaviors between $b = $ 0.1--1. However,
FFA becomes significant when $\dot{M}_w$ becomes very large at $b \lesssim
0.1$.  The observed data are not fitted well only by spectral break by FFA,
because the early rise of radio flux due to decreasing optical depth is more
rapid than SSA (see Weiler et al. 1986 for radio supernovae showing 
this feature), 
and it does not fit the observed slow rise of radio flux at
1.43 GHz. As a result, $\dot{M}_w$ cannot increase significantly with
decreasing $b$ at $b \lesssim 0.1$, and SSA frequency is always higher than
FFA for the best fit models (see Fig. \ref{fig:nu}). Because of this
constraint, the $\chi^2$ rapidly increase with decreasing $b$ smaller than
$\sim$ 0.1. The characteristic frequency $\nu_m$ is much smaller than GHz for
the low $\gamma_m$ value, but it becomes close to GHz for the high $\gamma_m$
value. This provides another freedom to the modeling of the observed
spectrum, and hence slightly better fits.

\subsection{Future emission by interaction with wind bubble or ISM}
\label{section:radio-future}
When the jet or fast ejecta is interacting with CSM of stellar wind profile,
the synchrotron flux decreases with time even if it is not
decelerated. However, it should eventually enter to a region where the
density is rather uniform. It is either interstellar medium (ISM) that has
not been affected by the progenitor star, or uniform density region of CSM
such as the stellar wind bubble composed by the shocked wind between the
contact discontinuity with ISM and wind termination shock (Weaver et
al. 1977; Koo \& McKee 1992).  If the jet is not yet decelerated in such
uniform density region, the synchrotron radiation flux should be simply
proportional to the mass of swept-up CSM/ISM matter, and hence should rapidly
increase with time as $\propto t^3$, until the jet sweeps up CSM/ISM matter
of a comparable mass with the jet itself and deceleration starts.\footnote{It
should be noted that the flux does not increase in the case of stellar wind
external profile, as seen in the previous section, even if the total energy
of shocked electrons increases as $\propto t$.  This is because the magnetic
field and hence synchrotron emissivity decrease with time by the scaling
assumed between the magnetic energy density and shocked matter.}  (Note that
the cooling frequency of electrons is much higher than the radio bands, and
hence cooling does not affect the radio flux.)  In the jet hypothesis
considered here, this occurs at a time $t_{\rm dec} \sim 13.9 b_{-1}^{-1/3}
(n_{\rm ext}/\rm cm^{-3})^{-1/3}$ yr after the explosion.  Here we assumed
that the uniform density region is composed mostly by hydrogen, and $n_{\rm
ext}$ is the hydrogen number density.  Therefore, once the jet enters to the
uniform external medium, we expect to see a rapid increase of the radio flux
in a time scale of years.

\begin{figure}
\plotone{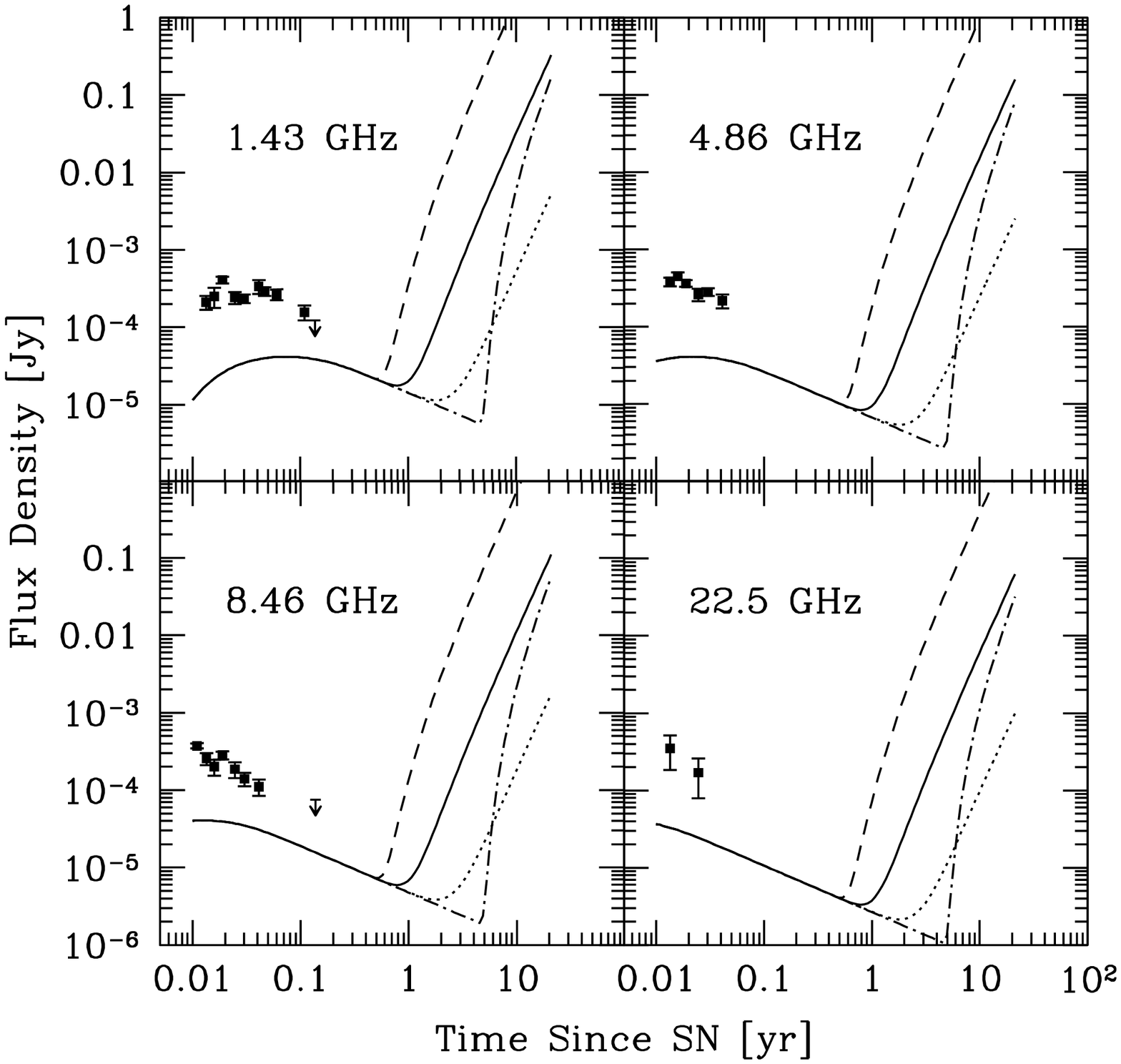}
\caption{Prediction of radio flux by CSM/ISM swept up by the 
jet in the future. 
Here we assumed that the observed radio flux (data points from
BKC02, error bars not including ISS modulation) 
is from CSM swept up by isotropic supernova ejecta
that is a different component from the jet. The parameters of
$(n_{\rm ext}{/\rm cm^{-3}}, r_{\rm ext}{\rm /cm})
= (1, 10^{17})$, $(0.1, 10^{17})$, $(10, 10^{17})$,
and $(1, 10^{18})$ are used for the solid, dotted, dashed, and
dot-dashed lines, respectively.
}
\label{fig:flux}
\end{figure}

The predicted radio flux by this process is shown in Fig. \ref{fig:flux},
where we used $b = 0.03$ (jet opening angular radius of $\sim 10^\circ$), $p
= 2.2$, $\epsilon_e = 0.05$, and $\epsilon_B = 0.01$, which are reasonable
values for GRB afterglows (Panaitescu \& Kumar 2001), as well as radio
supernovae.  We suppose the possibility (ii), and we adopted $\dot{M}_w = 5
\times 10^{-7} M_\odot$/yr, which is found by BKC02 by isotropic modeling.
The calculation is stopped at $t = t_{\rm dec}$, beyond which deceleration
must be taken into account.  The early radio flux by interaction between the
jet and wind-profile CSM is smaller than observed with these parameters, as
required for the possibility (ii).  Here we assumed that the density profile
becomes uniform beyond a radius $r_{\rm ext} = 10^{17}$ or $10^{18}$ cm, and
used three values of $n_{\rm ext}$ (0.1, 1, and 10 cm$^{-3}$ as typical
values for ISM as well as found in GRB afterglows). There is a discontinuity
in the density at $r_{\rm ext}$ depending on $\dot{M}_w$ and $n_{\rm ext}$,
but such discontinuity is also expected in reality for stellar wind bubbles
(e.g., Weaver et al. 1977).

The transition radius $r_{\rm ext}$ is not easy to estimate, but a
calculation of CSM density profile of presupernova Wolf-Rayet stars indicates
$r_{\rm ext} \sim 10^{17}$ cm (Ramirez-Ruiz et al. 2001).  Observations of
the prototype ring nebular NGC 6888 harboring a WR star indicate that the
wind termination shock is at $\lesssim$ 1--3 pc within the hot stellar wind
bubble (Wrigge, Wendker, \& Wisotzki 1994; Moore, Hester, \& Scowen 2000). On
the other hand, the radio light curves of SN 1998bw can be fitted with the
wind density profile up to $\sim$ a few $\times 10^{17}$ cm (Li \& Chevalier
1999).  Another indirect estimate of $r_{\rm ext}$ is possible from GRB
afterglow light curves, assuming that the environment of SN 2002ap is similar
to that of typical GRBs. At least some of GRB afterglow light curves can be
fit better by a model with homogeneous ambient density profile rather than
the wind profile (Chevalier \& Li 2000; Panaitescu \& Kumar 2002), and such
afterglow light curve seems to start at about 0.1--1 day after the burst, as
seen in the latest GRBs (021004, 021211) with very early optical detections
(Holland et al. 2003; Li et al. 2003; Fox et al. 2003). The observer's time
is related to the location of the shock of afterglows as $r = 3.0 \times
10^{17} (E_{\rm iso}/10^{52}\rm erg)^{1/4} (n_{\rm ext}/1\rm cm^{-3})^{-1/4}
(t_{\rm obs}/\rm 0.1 \ d)^{1/4}$ cm (e.g., Blandford \& McKee 1976), where
$E_{\rm iso}$ is the isotropic equivalent energy of GRBs, and beyond this
distance the density profile around GRBs seems uniform (see also Holland et
al. 2003).  Therefore, a plausible scale of $r_{\rm ext}$ is $\sim
10^{17}$--$10^{18}$ cm.

It should also be noted that the speed of isotropic supernova ejecta
producing the observed radio flux is inferred to be $\sim 0.3c$ (BKC02),
which is comparable or higher than the inferred jet speed of $0.23c$. Then
the isotropic ejecta producing the observed radio flux may be propagating
faster than the jet. In this case the jet is not directly interacting with
CSM and early radio flux prediction in the unshocked stellar wind region is
not valid. However, the isotropic supernova ejecta with a speed higher than
$0.3c$ is probably a tiny part of the total kinetic energy of the supernova,
and hence it will start to be decelerated much earlier than the jet. Then
eventually the jet will become the fastest and most distant component among
the supernova ejecta, and the radio flux prediction in Fig. \ref{fig:flux} is
valid in such a later epoch.

As expected, the radio flux shows a rapid increase at around $t_{\rm ext}
\equiv r_{\rm ext}/(\beta c) = 4.6 (r_{\rm ext}/10^{18} \ {\rm cm}) (\beta
/0.23)^{-1}$ yr.  The radio flux would become as strong as $\sim 0.1$ Jy.
This is not surprising when it is compared with radio GRB afterglows, which
have a similar energy scale but much higher expanding speed; typical GRB
afterglows would have radio flux of $\sim$ Jy for years after the burst, if
it is placed at a distance of SN 2002ap (Totani \& Panaitescu 2001).  
The expected radio flux for SN 2002ap in the near future
is strong enough to detect by a long-term monitoring.  Furthermore, jets
expanding to two opposite directions (both having $\theta_{\rm obs} \sim
90^\circ$) will have an angular separation of $\theta_{\rm sep} = 20 (t/{\rm
5 \ yr}) (\beta/0.23)$ mas, which is easy to spatially resolve by VLBI
observations, as proven for SN 1993J (Bartel et al. 1994; Marcaide et
al. 1995, 1997).  The jet directions should be perpendicular to the observed
polarization angle, giving a decisive proof for the jet hypothesis if it is
actually detected.  On the other hand, non-detection in next tens of years
would not necessarily exclude the jet hypothesis; if the density of the
uniform region is low ($n_{\rm ext} \lesssim 0.1 \ \rm cm^{-3})$ and $r_{\rm
ext}$ is large, it would take a long time ($\gtrsim$ a few tens of years) until
the radio emission becomes detectable again.

\section{Discussion}
\label{section:discussion}
\subsection{Implications for the GRB-SN Connection}
The jet inferred from SN 2002ap, having a similar energy scale to GRB jets,
should still be regarded as a hypothesis, and we should not over interpret it
at this time. However, since we have shown that the jet is physically
possible and it can be tested in the future, it is interesting to think what
the jet would mean and how it would fit 
to the other observational facts obtained
so far concerning the GRB-SN connection. Here we suggest a plausible picture
for the GRB-SN connection putting together 
the jet from SN 2002ap and other observational results,
assuming that the jet of SN 2002ap is real.

The similarity of the jet energy scale indicates that the SN 2002ap jet is
produced by the same mechanism as that for the cosmologically distant
GRBs. Then it may be called as a ``failed'' GRB, or another possibility is an
off-axis GRB, as mentioned in \S \ref{section:intro}.  Fortunately, we can
reject the latter possibility by radio observations made so far. If an
off-axis GRB is the case, then we expect even earlier re-appearance of radio
emission than discussed in \S \ref{section:radio-future}, which should show
even faster expansion with a velocity of $\sim c$, than the jet having a
velocity of $\sim 0.23c$. In fact, it is exactly what is called as ``orphan
afterglows'': a GRB whose jet direction is away from an observer, which can
be seen only by late time, less collimated afterglow emission. [For
discussions on detectability of radio emission from nearby GRB remnants, see
also Paczy\'nski (2001).]  As shown in Totani \& Panaitescu (2002), typical
GRB afterglows viewed at a large angle from the jet center at day $\sim$100
should have flux of $\sim$ 10--100 Jy, $V \sim$ 16--18, and $\sim (1$--$10)
\times 10^{-12} \ \rm erg \ cm^{-2} s^{-1}$, in radio (5GHz), optical, and
X-ray (1 keV) bands, respectively, at the distance to SN 2002ap.  These
fluxes are too faint to detect if it is placed at cosmological distances, but
thanks to the close distance to SN 2002ap, we should easily detect an
orphan afterglow of SN 2002ap. Then, the latest radio data of SN 2002ap about
200 days after the explosion, that is still at $\sim$ 0.1 mJy level showing
no evidence of flaring up (Berger et al. 2003), already exclude the off-axis
GRB possibility.

As discussed in \S \ref{section:nickel},
decay gamma-ray of $^{56}$Ni is the most likely ionization process of the jet
from SN 2002ap, but the required amount of $^{56}$Ni in the jet and/or
outside the photosphere is difficult to explain only by hydrodynamical
instability. Hence, this indicates that the jet is formed and ejected from the
central region of the core collapse. On the other hand, we do not expect
sufficient amount of $^{56}$Ni for ionization, if the jets are formed at
outer envelope of the star, such as trans-relativistic acceleration of shock
wave when it passes through the steep density gradient of stellar surface, as
studied by Matzner \& McKee (1999) and Tan, Matzner, \& McKee
(2001).\footnote{Following Tan et al. (2001), we use ``hydrodynamical shock
acceleration'' for this phenomenon, which must not be confused with the Fermi
acceleration of cosmic-ray particles.}

On the other hand, considering the similarity between SN 2002ap and 1998bw /
GRB 980425, we may also want to establish a consistent picture including GRB
980425 which is peculiar in a few points compared with GRBs found at
cosmological distances. GRB 980425 has soft spectrum and smooth temporal
profile (such as long spectral lag), and its total isotropic-equivalent
energy is about $10^4$ times smaller than typical GRBs at cosmological
distances (Bloom et al. 1998; Norris et al. 2000). One possible
interpretation of these properties is that this is a typical GRB but we
observed this event at a slightly off-axis direction from the jet center
(Nakamura 1999; Ioka \& Nakamura 2001; Salmonson 2001; Granot et
al. 2002). However, if GRB 980425 is a normal but off-axis luminous GRB, we
expect orphan afterglow radiation after it became
mildly-relativistic. Although optical afterglow radiation may be hidden by
the brightness of SN 1998bw (Granot et al. 2002), radio afterglow flux
expected for typical GRBs at a distance to SN 1998bw is 0.4--4 Jy at $t \sim
$ 100 days (Totani \& Panaitescu 2002), which is still much larger than
actually observed ($\sim$ 10 mJy at day 100), even though SN 1998bw was
much brighter than SN 2002ap. 
In addition to this, Norris (2002) identified a subclass
of GRBs including GRB 980425, whose members have low luminosity, long
spectral lags, and soft spectrum.  These low-luminosity GRBs tend to have
just a few wide pulses while nearly all high-luminosity GRBs have many,
narrow pulses, indicating that only the viewing angle effect cannot explain
the difference between the low- and high-luminosity GRBs.

Here we note our result that the radio emission from SN 2002ap cannot be
strongly collimated. Polarization was not detected for the strong radio
emission of SN 1998bw, indicating that the radio emission is produced by
isotropic ejecta (Kulkarni et al. 1998). The high velocity shell responsible
for radio flux of SN 1998bw shows an evidence of deceleration (Li \&
Chevalier 1999), and hence the kinetic energy of the shell should not be much
larger than $\sim 10^{49}$ erg. GRB 980425 associated with SN 1998bw has
isotropic equivalent energy of about $\sim 10^{48}$ erg, which is very
different from the standard energy scale of cosmologically distant GRBs, but
is close to that of isotropic radio emitting shell.

Tan et al. (2001) have shown that the peculiar GRB 980425 can be produced by
the isotropic fast shell produced by the hydrodynamical shock acceleration,
which is the same shell producing radio emission.  The smooth temporal
profile of GRB 980425 may also be expected by isotropic shock acceleration
process. Even if explosion is highly asymmetric at the
center, a two-dimensional simulation by Maeda et al. (2002) indicates that
the outer low-density region is rather isotropic.  The efficiency of the
shock acceleration sensitively depends on the outer density profile of
stellar envelope (Tan et al. 2001), and it may not be surprising even if SN
1998bw and SN 2002ap produced very different velocities and energies of fast
moving isotropic ejecta.

Then, the apparent discreteness between GRB 980425 and other GRBs may be
understood by two distinct driving processes of GRBs: cosmologically distant
GRBs are strongly collimated jets with an energy scale of $\sim 10^{50-51}$
erg, which is produced along with ejection of $^{56}$Ni outside the
photosphere by the central activity of core collapses (refereed as type 1 for
convenience), while GRB 980425 is produced by much less energetic, isotropic
fast ejecta, which could be produced by hydrodynamic shock acceleration at
outer layer of exploding stars (type 2). There is a possibility that SN
1998bw was successful also as off-axis type 1 GRB, but this possibility is
not favored by the no detection of radio orphan afterglow showing an energy
scale of $\sim 10^{51}$ erg for $\sim$ 300 days, as mentioned above.

Other past events of type Ic supernovae may also have produced energetic jets
with a velocity of $\sim 0.2c$, like SN 2002ap. Such jets, if exist, may now
be emitting strong radio emission after sweeping up enough amount of ISM/CSM,
showing jet-like morphology. We note that the time scale of emergence in
radio wavelength could be much longer for the case of a failed GRB with
low-speed jets, than considered in previous publications for off-axis GRBs or
GRB remnants (Paczy\'nski 2001).  On the other hand, it would be shorter than
the time scale of establishment of normal radio supernova remnants,
corresponding to the transition from the free expansion to Sedov-Taylor phase
($\sim 100$ yr). Re-examination and new follow-up of such past type Ic events
are encouraged.


The picture of GRB-SN connection presented here predicts that all type 1 GRBs
should be associated with energetic type Ib/Ic supernovae (as confirmed by SN
2003dh / GRB 980329 after the submission of this paper), but we expect a
variety of supernova luminosity that may not be correlated with GRB
luminosity or the jet energy.  The supernova luminosity is determined by the
amount of synthesized $^{56}$Ni within the photosphere where the radioactive
energy is converted into optical photons. As we suggested, the jet formation
activity may eject a significant amount of $^{56}$Ni well outside the
photosphere, where $^{56}$Ni cannot contribute to the optical luminosity.
This is not inconsistent with the results of search for supernova evidence in
GRB afterglows (Bloom et al. 1999; Galama et al. 2000; Reichart 2001; Bloom
et al. 2002; Garnavich et al.  2003; Price et al. 2003).

\subsection{On the X-ray line features in GRB afterglows}
\label{section:X-ray-line}
Emission line features of iron (or nickel) and multiple-alpha nuclei (Mg, Si,
S, Ar, Ca, etc.)  are often found in X-ray GRB afterglows on a time scale of
$\sim$ a day (Piro et al. 1999, 2000; Yoshida et al. 1999; Antonelli et
al. 2000; Reeves et al. 2002; Butler et al. 2003).  Theoretical explanations
mostly fall into the two categories: (1) geometry-dominated (GD) scenario
where the time scale of $\sim$ a day is attributed to the photon propagation
time, and this scenario needs a supernova explosion that occurred weeks prior
to a GRB (the supranova model, Vietri et al. 2001), dense preburst
circumstellar material (Weth et al. 2000; Kotake \& Nagataki 2001) or a
distant reflector of GRB/afterglow emission such as a e$^\pm$-pair screen
(Kumar \& Narayan 2003); and (2) engine-dominated (ED) model where a
long-lived energy source left over in the center of the star after the GRB is
ionizing the matter around it (Rees \& M\'esz\'aros
2001).  Most of these explanations assume that ionization process is
photoionization by GRB/afterglow radiation or long-lived remnants, while an
alternative is shock heating around the center (B\"ottcher 2000). However,
all these scenarios have one or more problems (Lazzati, Ramrez-Ruiz, \& Rees
2002; Kumar \& Narayan 2003), and other possible explanations are still worth
seeking for.

Here we propose that the ionization by radioactivity of decaying $^{56}$Ni in
mildly optically thin region, which has been suggested as the ionization
mechanism for the SN 2002ap jet, can be considered as a new explanation for
the X-ray line features of afterglows. First let us check the energetics.
The luminosity of decaying energy is given by $\sim 1.0 \times 10^{44}
(M_{\rm Ni}/M_\odot)$ erg/s within $\sim t_{\rm Ni} = $ 8.1d after the
explosion. We need several solar mass of $^{56}$Ni to explain the
luminosities of X-ray line features; this is admittedly large compared
with typical supernovae, and here we have assumed 100\% efficiency for
the energy conversion from radioactive decay into line photons, which 
seems rather extreme. 
However, we know that SN 1998bw produced about $\sim
1 M_\odot$ of nickel, and it seems not very unlikely that more massive and
energetic supernova (or hypernova) events produce even more nickel than SN
1998bw and eject it out to relatively optically thin region, via the
process of jet formation for type 1 GRBs. If radioactive gamma-rays
lose their energy dominantly by ionizing neaby ions, efficiency of
line production could be close to the unity in some preferred situations.
If all $^{56}$Ni ions are fully ionized, the radioactive decay by electron
capture is impossible (McLaughlin \& Wijers 2002), but repeated decays and
ionizations would be possible if recombination rate is sufficiently high.
This is a complicated process, and clearly more careful and quantitative
study is required to support this speculation.

This scenario can be considered as a new type of ED explanations that do not
need a supernova prior to a GRB, but has one important difference from
previous ones; this scenario does not require strong radiation source in
X-ray band for ionization, but high energy electrons produced by scattering
of decay gamma-rays directly ionize ions, and hence we might expect high
equivalent width with weak continuum radiation. In this way the major problem
of ED scenarios, i.e., that we should have directly observed stronger
ionizing radiation or bremsstrahlung continuum of shock-heated matter than
actually observed (Lazzati et al. 2002; Kumar \& Narayan 2003), might be
avoided. 

The time scale of line production should be determined by evolution of
physical conditions by expansion, such as optical depth, density, and
recombination rates. Note again that radioactive ionization should be
efficient only in a small range of optical depth where it is mildly optically
thin, and then we expect a shorter time scale of line production than typical
peak of supernova light curve that is determined by diffusion of radiation
within the photosphere, and it might become even shorter than the time scale
of $^{56}$Ni decay.


\section{Summary and Conclusions}
\label{section:conclusion}
We have shown that the jet hypothesis proposed by Kawabata et al. (2002) for
SN 2002ap based on their spectropolarimetric observation at day $\sim 13$ is
physically possible and consistent with all observations. We estimated the
jet mass ($\sim 0.01 M_\odot$), direction (approximately perpendicular to the
observer), and energy ($\sim 5 \times 10^{50}$ erg) by a fully relativistic
treatment. The large jet energy does not contradict with the weak radio flux,
since the jet is almost freely expanding and jet material is not yet
shocked. The
total energy of electrons inferred from synchrotron radio flux, $\sim
10^{45}$ erg, only reflects that of CSM swept-up and shocked by jet or
high-speed ejecta from SN 2002ap. The jet becomes optically thin on a time
scale of 5--10 days, and weak continuum polarization found by the earlier VLT
observation at day $\sim 5$ can be explained by high optical thickness of the
jet to electron scattering.

The jet must be substantially ionized to reproduce the redshifted polarized
continuum.  Rapid adiabatic loss of the jet internal energy should have
cooled down the jet material below the temperature required for the jet to be
kept ionized, and hence external heating or photoionization source is
necessary.  We have shown that photoionization is quite unlikely by any
sources, and the most likely ionization process is heating by decaying
$^{56}$Ni when the jet is mildly optically thin. Still, ionization is not
easy; the jet must have a mass fraction of order unity of
$^{56}$Ni, or a larger amount of $^{56}$Ni outside the photosphere that is at
least comparable with that within the photosphere ($\sim 0.07 M_\odot$),
which is difficult to explain simply by hydrostatic instability. This result
indicates that the jet must be formed and ejected from the central region of
the core collapse.

We examined whether the observed radio emission can be explained by CSM
swept-up by the jets, and we found that the radio data favor isotropic
ejecta. Jets with $b \equiv (\Omega_{\rm jet} / 4 \pi)
\sim 0.1$ is marginally possible to explain the radio
data, and stronger collimation is excluded as the origin of the observed
radio emission. If the jet is strongly collimated, the radio emission
must be from isotropic fast ejecta that is a different component from the
jet.

The jet should be regarded still a hypothesis, but we predict that, if the
jet hypothesis is the case, the jet would eventually pass through the
unshocked presupernova wind region ($\rho \propto r^{-2}$), and enter to a
region where the CSM/ISM density is rather uniform.  The jet mass is large
enough not to be decelerated until it sweeps up a comparable mass of ISM/CSM
in a few to $\sim$10 years. Then we expect a rapid increase of radio flux
that should easily be detected by future long-term radio monitoring, with
reasonable parameters for ISM/CSM density and profile.  Fortunately, such
radio emission can be resolved spatially by a VLBI observation, and the
morphology and jet direction relative to the observed polarization angle 
would give a clear proof of the jet hypothesis.

Although it is a speculation that would become effective when the jet
hypothesis is confirmed by future observations, we discussed how the inferred
jet would fit to the our knowledge about the GRB-SN connection, taking into
account various observational facts obtained in the past. To explain all the
observations consistently, we suggest two distinct classes of GRBs by
different formation processes but from similar core collapse events;
cosmologically distant GRBs are produced by strongly collimated jets having
an energy scale of $\sim 10^{50-51}$ erg, which are produced and ejected from
the central region of the core collapse.  SN 2002ap jet can be considered as
a failed GRB of this class, with much larger baryon load than typical
successful GRBs. $^{56}$Ni is ejected from the center
along with the jet, which is responsible for the ionization of the SN 2002ap
jet. Such $^{56}$Ni may also explain the X-ray line features often found in
GRB afterglows, depending on the amount of $^{56}$Ni and physical conditions,
although more quantitative study is required to verify this possibility. 
On the other hand, the
peculiar GRB 980425 and radio emission from SN 1998bw and SN 2002ap that seem
isotropic may be produced by isotropic and much less energetic ejecta, which
could be formed by the hydrodynamical shock acceleration at the surface of an
exploding star.

The author would like to thank M. Iye, K.S. Kawabata, K. Maeda, K. Nomoto,
B. Paczy\'nski, J. C. Tan, and J.D. Akita
for stimulating information and discussions. The author has
been financially supported in part by the JSPS Fellowship for Research
Abroad.


\epsscale{0.7}

\epsscale{0.5}

\begin{references}

\reference{} Antonelli, L.A. et al. 2000, ApJ, 545, L39

\reference{} Arnett, W.D. 1979, ApJ, 230, L37

\reference{} Bartel, N. et al. 1994, Nature, 368, 610

\reference{} Berger, E., Kulkarni, S.R., \& Chevalier, R.A. 2002, ApJ, 577, L5
(BKC02)

\reference{}  Berger, E., Kulkarni, S.R., Frail, D.A., \& Soderberg, A.M.
2003, astro-ph/0307228

\reference{} Blandford, R.D., McKee, C.F. 1976, Phys. Fluids, 19, 1130

\reference{} Bloom, J.S., Kulkarni, S.R., Harrison, F., Prince, T.,
Phinney, E.S., \& Frail, D.A. 1998, ApJ, 506, L105

\reference{} Bloom, J.S. et al. 1999, Nature, 401, 453

\reference{} Bloom, J.S. et al. 2002, ApJ, 572, L45

\reference{} B\"ottcher, M. 2000, ApJ, 539, 102

\reference{} Butler, N.R. et al. 2003, submitted to ApJ, astro-ph/0303539

\reference{} Chevalier, R.A. \& Fransson, C. 2001, in
``Supernovae and Gamma-Ray Bursts'' ed. K. W. Weiler (Springer-Verlag)
(astro-ph/0110060)

\reference{} Chevalier, R.A. \&  Li, Z.-Y. 2000, ApJ, 536, 195

\reference{} Frail, D.A. et al. 2001, ApJ, 562, L55

\reference{} Fransson, C. 1982, A\&A, 111, 140


\reference{} Galama, T.J. et al. 2000, ApJ, 536, 185

\reference{} Gal-Yam, A., Ofek, E.O., \& Shemmer, O. 2002, MNRAS, 332, L73

\reference{} Garc\'ia-Segura, G., Mac Low, M.-M., \& Langer, N. 1996,
A\&A, 305, 229 

\reference{} Garc\'ia-Segura, G., Langer, N., \& Mac Low, M.-M. 1996,
A\&A, 316, 133 

\reference{} Garnavich, P.M. et al. 2003, ApJ, 582, 924

\reference{} Goodman, J. 1986, ApJ, 308, L47

\reference{} Graham, J.R. 1988, ApJ, 335, L53

\reference{} Granot, J., Panaitescu, A., Kumar, P., \& Woosley, S.E.
2002, ApJ, 570, L61

\reference{} H\"oflich, P., Wheeler, J.C., \& Wang, L. 1999, ApJ, 521, 179

\reference{} Hjorth, J. et al. 2003, Nature, 423, 847 

\reference{} Holland, S.T. et al. 2003, AJ, in press (astro-ph/0211094)

\reference{} Huang, Y.F., Dai, Z.G., \& Lu, T. 2002, MNRAS, 332, 735

\reference{} Hurley, K. et al. 2002, GCN Circ. 1252

\reference{} Ioka, K. \& Nakamura, T. 2001, ApJ, 554, L163

\reference{} Iwamoto, K. et al. 1998, Nature, 395, 672

\reference{} Kawabata, K.S. et al. 2002, ApJ, 580, L39

\reference{} Kinugasa, K. et al. 2002, ApJ, 577, L97

\reference{} Koo, B.-C. \& McKee, C.F. 1992, ApJ, 388, 93

\reference{} Kotake, K. \& Nagataki, S. 2001, PASJ, 53, 579

\reference{} Kulkarni, S.R. et al. 1998, Nature, 395, 663

\reference{} Kumar, P. \& Narayan, R. 2003, ApJ, 584, 895

\reference{} Lazzati, D., Ramirez-Ruiz, E., \& Rees, M.J. 2002, 
ApJ, 572, L57

\reference{} Leonard, D.C., Filippenko, A.V., Chornock, R., \& Foley,
R.J. 2002, PASP, 114, 1333

\reference{} Li, Z.-Y. \& Chevalier, R.A. 1999, ApJ, 526, 716

\reference{} Li, W., Filippenko, A.V., Chornock, R., Jha, S. 2003,
ApJ Lett. in press, astro-ph/0302136

\reference{} Longair, M.S. 1992, High Energy Astrophysics vol. 1
(Cambridge; Cambridge)

\reference{} Lotz, W. 1967, ApJS, 14, 207

\reference{} Lucy, L.B. 1991, ApJ, 383, 308

\reference{} Maeda, K., Nakamura, T., Nomoto, K., Mazzali, P.A.,
Patat, F. \& Hachisu, I. 2002, ApJ, 565, 405

\reference{} Marcaide, J.M. et al. 1995, Sci, 270, 1475

\reference{} Marcaide, J.M. et al. 1997, ApJ, 486, L31

\reference{} Matzner, C.D. \& McKee, C.F. 1999, ApJ, 510, 379

\reference{} Mazzali, P.A. et al. 2002, ApJ, 572, L61

\reference{} McCray, R. 1983, Proc. 18th IAU General Assembly, Highlights
of Astronomy, Vol. 6. Reidel, Dordrecht, p565

\reference{} McLaughlin, G.C., Wijers, R.A.M.J., Brown, G.E., \&
Bethe, H.A. 2002, ApJ, 567, 454

\reference{} McLaughlin, G.C. \& Wijers, R.A.M.J. 2002, ApJ, 580, 1017

\reference{} M\'esz\'aros, P., Laguna, P., \& Rees, M.J. 1993, ApJ, 415, 181


\reference{} Moore, B.D., Hester, J.J., \& Scowen, P.A. 2000,
AJ, 119, 2991

\reference{} Nagataki, S. 2000, ApJS, 127, 141

\reference{} Nahar, S.N. \& Pradhan, A.K. 1997, ApJS, 111, 339

\reference{} Nahar, S.N. 1999, ApJS, 120, 131

\reference{} Nakamura, T. 1999, ApJ, 522, L101

\reference{} Nakamura, T., Mazzali, P.A., Nomoto, K., \& Iwamoto, K.
2001, ApJ, 550, 991

\reference{} Nomoto, K., Filippenko, A.V., \& Shigeyama, T. 1990,
A\&A 240, L1

\reference{} Norris, J.P., Marani, G.F., \& Bonnell, J.T. 2000,
ApJ, 534, 248

\reference{} Norris, J.P. 2002, ApJ, 579, 386

\reference{} Osterbrock, D.E., 1989, Astrophysics of Gaseous Nebulae
(California; University Science Books)

\reference{} Paczy\'nski, B. 1986, ApJ, 308, L43

\reference{} Paczy\'nski, B. 2001, Acta. Astron., 51, 1

\reference{} Panaitescu, A. \& Kumar, P. 2001, ApJ, 560, L49

\reference{} Panaitescu, A. \& Kumar, P. 2002, ApJ, 571, 779

\reference{} Patat, F. et al. 2001, ApJ, 555, 900

\reference{} Piro, L. et al. 1999, ApJ, 514, L73

\reference{} Piro, L. et al. 2000, Science, 290, 955

\reference{} Press, W.H., Teukolsky, S.A., Vetterling, W.T., \&
Flannery, B.P. 1992, Numerical Recipes (Cambridge; Cambridge)

\reference{} Price, P.A. et al. 2003, ApJ, 584, 931

\reference{} Ramirez-Ruiz, E., Dray, L.M., Madau, P., \& Tout, C.A.
2001, MNRAS, 327, 829

\reference{} Reeves, J.N. et al. 2002, Nature, 416, 512

\reference{} Rees, M.J. \& M\'esz\'aros, P. 2000, ApJ, 545, L73

\reference{} Reichart, D.E. 2001, ApJ, 554, 643

\reference{} Rybicki, G.B. \& Lightman, A.P. 1979, Radiative Processes
in Astrophysics (New York; Wiley)

\reference{} Salmonson, J.D. 2001, ApJ, 546, L29

\reference{} Schaere, D. \& de Koter, A. 1997, A\&A, 322, 598

\reference{} Sharina, M.E., Karachentsev, I.D., \& Tikhonov, N.A. 1996,
A\&AS, 119, 499

\reference{} Shigeyama, T., Nomoto, K., Tsujimoto, T., \& Hashimoto, M.
1990, ApJ, 361, L23

\reference{} Smartt, S.J. et al. 2002, ApJ, 572, L147

\reference{} Smartt, S.J. \& Meikle, P. 2002, IAUC 7822

\reference{} Sohn, Y.-J. \& Davidge, T.J. 1996, AJ, 111, 2280

\reference{} Soria, R. \& Kong, A.K.H. 2002, ApJ, 572, L33

\reference{} Stanek, K.Z. et al. 2003, ApJ, 591, L17

\reference{} Sutaria, F.K., Chandra, P., Bhatnagar, S., \& Ray, A.
2003, A\&A, 397, 1011

\reference{} Tan, J.C., Matzner, C.D., \& McKee, C.F. 2001, ApJ, 551, 946

\reference{} Totani, T. \& Panaitescu, A. 2002, ApJ, 576, 120

\reference{} Vietri, M., Ghisellini, G., Lazzati, D., Fiore, F.,
\& Stella, L. 2001, ApJ, 550, L43

\reference{} Wang, L. et al. 2002, preprint submitted to ApJ, astro-ph/0206386

\reference{} Weaver, R., McCray, R., Castor, J., Shapiro, P., \&
Moore, R. 1977, ApJ, 218, 377

\reference{} Weiler, K.W., Sramek, R.A., Panagia, N., van der Hulst,
J.M., \& Salvati, M. 1986, ApJ, 301, 790

\reference{} Weth, C., M\'esz\'aros, P., Kallman, T., \& Rees, M.J.
2000, ApJ, 534, 581

\reference{} Wrigge, M., Wendker, H.J., \& Wisotzki, L. 1994, A\&A,
286, 219

\reference{} Woosley, S.E., Pinto, P.A., \& Hartmann, D. 1989, ApJ, 346, 395

\reference{} Woosley, S.E., Eastman, R.G., \&
Schmidt, B.P. 1999, ApJ, 516, 788

\reference{} Yoshida, A. et al. 1999, A\&AS, 138, 433

\end{references}
\end{document}